\documentclass[aps,pra,reprint,noshowpacs,nofootinbib,superscriptaddress]{revtex4-1}  % for double-spaced preprint
\usepackage{graphicx} 	% needed for figures
\usepackage{dcolumn}  	% needed for some tables
\usepackage{bm}              	% for math
\usepackage{amssymb}   	% for math
\usepackage{amsmath}   	% for math
\usepackage{braket}         %for brakets
\usepackage{hyperref}

\def\be{\begin{equation}}
\def\ee{\end{equation}}
\def\bes{\begin{eqnarray}}
\def\ees{\end{eqnarray}}

\begin{document}
\widetext
%\leftline{updated on \today}
%\leftline{Rough Draft}

\title{Demonstration of Bayesian quantum game on an ion trap quantum computer}
 
\author{Neal Solmeyer}\let\thempfn\relax\footnotetext[0]{This work was conducted while N.\ Solmeyer was at the Army Research Laboratory. The author's affiliation with the MITRE Corporation is provided for identification purposes only, and is not intended to convey or imply MITRE's concurrence with, or the support for, the positions, opinions or viewpoints expressed by the author. Approved for Public Release;
Distribution Unlimited. Case Number 17-4775.}\email{nsolmeyer@mitre.org} 
\affiliation{
	The MITRE Corporation
	 7596 Colshire Drive, McLean, VA, 22102-7539, USA.}
\author{Norbert M.\ Linke}
\affiliation{Joint Quantum Institute, Department of Physics,
	and Joint Center for Quantum Information and Computer Science,
	University of Maryland, College Park, MD 20742, USA.}
\author{Caroline Figgatt}
\affiliation{Joint Quantum Institute, Department of Physics,
	and Joint Center for Quantum Information and Computer Science,
	University of Maryland, College Park, MD 20742, USA.}
\author{Kevin A.\ Landsman}
\affiliation{Joint Quantum Institute, Department of Physics,
	and Joint Center for Quantum Information and Computer Science,
	University of Maryland, College Park, MD 20742, USA.}
\author{Radhakrishnan Balu}
\email{radhakrishnan.balu.civ@mail.mil}
\affiliation{
	Computer and Information Sciences Directorate, Army Research Laboratory, Adelphi, MD, 21005-5069, USA. }
\affiliation{Computer Science and Electrical Engineering,
	University of Maryland Baltimore County,
	1000 Hilltop Circle, Baltimore, MD 21250, USA.}
\email{radbalu1@umbc.edu}

\author{George Siopsis}
\email{siopsis@tennessee.edu}
\affiliation{Department of Physics and Astronomy,
	The University of Tennessee, Knoxville, TN 37996-1200, USA.}
\author{Christopher Monroe}
\affiliation{Joint Quantum Institute, Department of Physics,
and Joint Center for Quantum Information and Computer Science,
University of Maryland, College Park, MD 20742, USA.} 
\affiliation{IonQ Inc., College Park, MD, 20740, USA.}
\date{\today}

\begin{abstract} 
We demonstrate a Bayesian quantum game on an ion trap quantum computer with five qubits. The players share an entangled pair of qubits and perform rotations on their qubit as the strategy choice. Two five-qubit circuits are sufficient to run all 16 possible strategy choice sets in a game with four possible strategies. The data are then parsed into player types randomly in order to combine them classically into a Bayesian framework. We exhaustively compute the possible strategies of the game so that the experimental data can be used to solve for the Nash equilibria of the game directly. Then we compare the payoff at the Nash equilibria and location of phase-change-like transitions obtained from the experimental data to the theory, and study how it changes as a function of the amount of entanglement. 
\end{abstract}

\maketitle

\section{introduction}
Game theory, originally developed in the 1940s and 1950s  \cite{Math,Nash1,Nash2}, has since been the source of important contributions in fields such as economics \cite{Econ}, political science \cite{Pol}, biology \cite{Bio}, and computer science \cite{CS}. The addition of quantum information concepts into games led to the formalization of quantum games \cite{Meyer1999,Eisert1999}. Since their introduction, quantum games have been studied in a variety of contexts. With the growing prevalence of quantum computers and quantum networks,  quantum games emerge as strong candidates for real world applications in quantum security protocols \cite{Maitra2015}, distributed quantum computing algorithms \cite{Li2009}, or improving the efficiency of classical network routing algorithms \cite{Zabaleta2017}.

In contrast with many interesting quantum computing algorithms, quantum games can be demonstrated with small numbers of qubits, making them an attractive application for early demonstrations on quantum computers. There have been several experimental demonstrations of quantum games using NMR quantum computers \cite{Du2002,Mitra2007} and various linear optical quantum computing schemes \cite{Prevedel2007,Zhang2008,Altepeter2009,Schmid2010,Zu2012,Pinheiro2013b,Balthazar2015}.  In these demonstrations, the circuit equivalent of the game was executed with the strategy choices of the Nash equilibrium, as determined by theoretical analysis, and compared the expected payoff at Nash equilibrium. This amounts to a type of benchmarking of the performance of the quantum computer under the framework of game theory. 

One might argue that a true demonstration of a quantum game would actually involve real players, either humans or computers, playing the game on true quantum hardware. With real players playing on a classically simulated quantum computer, people may actually play quantum games more rationally than they play classical games, even if they have no prior training in quantum mechanics \cite{Chen2006,Chen2008}. This may be due to the fact that people may have fewer preconceived notions about the quantum strategies, and are thus more likely to simply play for the highest payoff. Though this may have interesting implications for potential real world applications of quantum games, there is a gap between the potential uses of quantum games and the availability of quantum hardware. 

This work aims to partially address that gap by performing a more complete demonstration of a game on a scalable quantum architecture that can be applied to more complex game scenarios in the future. Ion trap based systems are promising candidates for quantum computers, which is a prerequisite for quantum games. Further, ion-trap architectures are also promising for quantum networking \cite{Blinov2004}, where nodes are remotely located and entangled, which may also be a requirement for quantum game applications that require the players to be remotely located. 

We realize a game with incomplete information, i.e. a Bayesian game \cite{Harsanyi1967}. The amount of incomplete information is determined by a probability distribution of different player types. Bayesian games are of interest because of their deep connection to Bell's inequalities \cite{Brunner2013}. The Bayesian game we analyze is not directly derived from a Bell's inequality, but is rather a Bayesian game formed by incorporating incomplete information classically into a quantum game. The motivation behind this approach is that incomplete information may often be a feature of any potential application of quantum games, not just games specifically designed to violate Bell's inequalities, and the interplay of classical probability and quantum statistics can lead to a rich structure \cite{Solmeyer2017}. 

This is the first experimental demonstration of a quantum game using a scalable architecture. Because of the sophistication of current ion-trap quantum computers compared to previously used quantum computers, a more extensive implementation is possible.  Using a novel parallelization scheme, we perform an exhaustive computation of all possible strategy choice combinations. This allows us to solve for the Nash equilibria of the game based only on the experimental outcomes. Therefore, we can look not only at the payoff at Nash equilibria, but also study where the equilibria occur and observe the effect of experimental noise on the phase-change-like behavior of quantum games. Such behavior occurs when one set of equilibria changes to another as the entanglement or amount of incomplete information is changed \cite{Du2003}.  We find that the deviation of the expected to measured payoff grows with the degree of entanglement in the game. We also find that the deviation between the experimental and theoretical location of the phase-change-like thresholds grows with entanglement. 

\section{Quantum game implementation}
We experimentally demonstrate a Bayesian game that displays several features worth exploring such as multiple simultaneous Nash equlibria and phase-change-like behavior as functions of both the Bayesian probability and the amount of entanglement.  A detailed theoretical analysis of this game can be found in ref.~\cite{Solmeyer2017}.  Being composed of two player games, the demonstration is relatively straightforward to implement experimentally, yet has a rich enough structure to observe several features of Bayesian games that are interesting from a game theoretical perspective.

\begin{figure}
\includegraphics[width=\columnwidth]{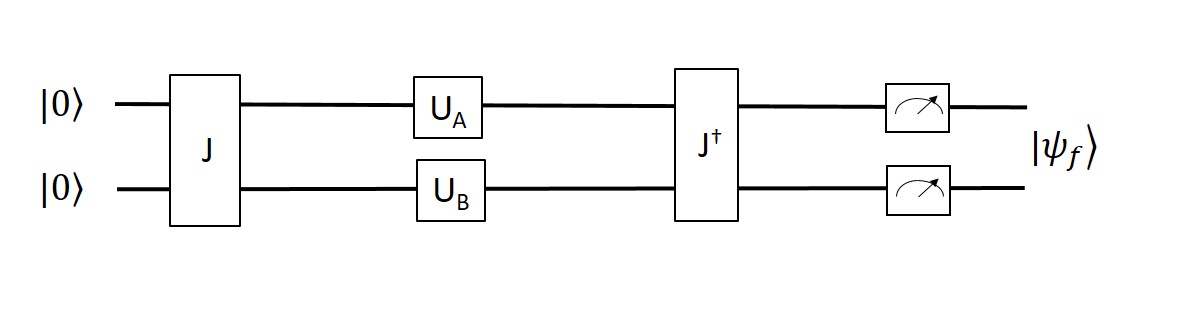}
\caption{\label{fig:circuit} Basic circuit of two player game. The qubits are entangled $J(\chi)$, the players apply their strategy choices $U_A$ and $U_B$, the qubits are unentangled $J^\dagger(\chi)$, and finally a measurement is performed. }
\end{figure}

First, we describe the two player game, which is used to construct the Bayesian game, see Fig. ~\ref{fig:circuit}. An entangling gate is applied to two qubits, initialized in the state $\ket{00}$, each qubit corresponding to one of  two players. The entangling operation is given by the general XX gate:

\begin{equation}
\begin{aligned}
 J(\chi) =& e^{i\chi X\otimes X} = \cos\chi + i X\otimes X \sin\chi \\
=&\begin{pmatrix}
\cos\chi & 0 &0& - i \sin\chi \\
0 &\cos\chi &- i \sin\chi&0 \\ 
0 & - i \sin\chi &\cos\chi&0 \\
- i \sin\chi & 0 &0&\cos\chi \\ 
\end{pmatrix},
\end{aligned}
\end{equation}
where $\chi \in [0,\pi/4]$. After the entangling operation, the players apply their strategy choices $U_A$ and $U_B$. Finally we will need the conjugate transpose of $J(\chi),  (J(\chi))^\dagger= J(-\chi)$. The final state for the two player game is given, in the basis $\ket{A, B}$, by:
\begin{equation}
\ket{\psi_{f}} = J^{\dagger}(\chi) \cdot (U_A \otimes U_{B1}) \cdot  J(\chi) \ket{00}. 
\label{eq:circuits}
\end{equation}

The payoff is then calculated based on the measurements of the qubits. If the outcome of the measurement is $\ket{0}$, this corresponds to one strategy choice in the classical game (the analogue of cooperation (C) in the prisoner's dilemma), whereas if the outcome is $\ket{1}$, this corresponds to the analogue of defection (D). The payoff for a given strategy choice set is determined by the payoff matrix for the game:
\begin{center}
\begin{tabular}{ l || c | r }
$A| B_1$ & $ \Ket{0}(C)$ &$\Ket{1}(D)$\\
\hline 
\hline
$\Ket{0}(C)$& $ (11,9)$ & $(1,10)$ \\
$\Ket{1}(D)$& $ (10,1)$ & $(6,6)$ \\

\end{tabular}

\medskip

\begin{tabular}{ l || c | r }
$A| B_2$ & $ \Ket{0}(C)$ &$\Ket{1}(D)$\\
\hline 
\hline
$\Ket{0}(C)$& $ (11,9)$ & $(1,6)$ \\
$\Ket{1}(D)$& $ (10,1)$ & $(6,0)$ 
\end{tabular}
\end{center}

The two payoff matrices are for the games versus the two different player types for player B. The rows and columns represent the outcomes (which are the strategy choices in classical games) of player A and B respectively, and the numbers $(\$^A,\$^B)$ are the payoffs for player $A$ and $B$. 

The game between $A$ and $B_1$ is the standard prisoner's dilemma, while in the game between $A$ and $B_2$, player $B_2$ believes player $A$ is the DA's brother, which gives player $A$ an advantage resulting in an asymmetric payoff between the players.

The payoff for player $A$ is given by the expectation value of the final state weighted by the elements of the payoff matrix: 
\begin{equation}
\langle\$^A \rangle = \sum_i | \braket{i | \psi_f} |^2 \$^A_i,
\label{eq:payoff}
\end{equation}
where the sum is over all four possible measurement outcomes of the two qubit system, and the  $ \$^A_i$ are the corresponding elements of the payoff matrix.

A game is also defined by the allowable strategy choices. We implement the game using four possible strategy choices, compared to the most commonly used set of three. The four-choice single player strategy set, which we label $U$, is given by the three Pauli matrices ($X, Y$, and $Z$) plus the identity ($I$). This choice of possible strategies bounds the results of a game with arbitrary continuous strategy choices \cite{Landsburg2011,Solmeyer2017b}. 

The Bayesian game consists of two players $A$ and $B$. Player $A$ is of one type, and player $B$ can be one of two types, $B_1$ or $B_2$. Player $A$ plays with either $B_1$ or $B_2$, with some probability $p$, which parametrizes the incomplete information held by player $A$ which is maximal when $p = 0.5$, and minimal when $p=0$ or $1$.  This game uses three qubits, one for each player type, and proceeds as follows. All three qubits are initialized to state $\ket{0}$. The entangling operation $J(\chi)$ is performed between either $A$ and $B_1$ or $A$ and $B_2$ probabilistically. Each player chooses a strategy $U_i$ from the set $U$ and applies it to their qubit. Then the appropriate unentangling operation $J^\dagger(\chi)$ is performed.  Finally, the three qubits are measured and the expected payoff for player $A$ is given by the weighted average of playing with $B_1$ and $B_2$:
\begin{equation}
\langle\$^A \rangle = \langle\$^A(A,B_1) \rangle (p)+ \langle\$^A(A,B_2) \rangle(1-p) 
\label{eq:payoffB}
\end{equation}
and the payoff for the $B$ players is given by $\langle \$^{B1}(A,B_1)\rangle$ and $\langle \$^{B2}(A,B_2)\rangle$. There is structure in the game as the amount of entanglement, $\chi$, is varied, and as the probability to play with either player, $p$, is varied. We demonstrate this by varying $\chi$ in the experiment, and varying $p$ in the analysis. To implement a Bayesian game between players $A$, $B_1$, and $B_2$, we run many versions of the two player circuit shown in Fig. \ref{fig:circuit}, and then combine them into a Bayesian game in the data analysis. This is similar to what has been done to produce initial states for quantum games which are a mixture of entangled states that exhibit quantum discord \cite{Zu2012}.

The demonstration requires running $4 \times 4 = 16$ two-qubit circuits for different strategy combinations, and to make efficient use of the hardware available, we employ a novel parallelization scheme. We compute the circuit for multiple strategy choices simultaneously by using auxiliary qubits in superposition.

\begin{figure}
\includegraphics[width=\columnwidth]{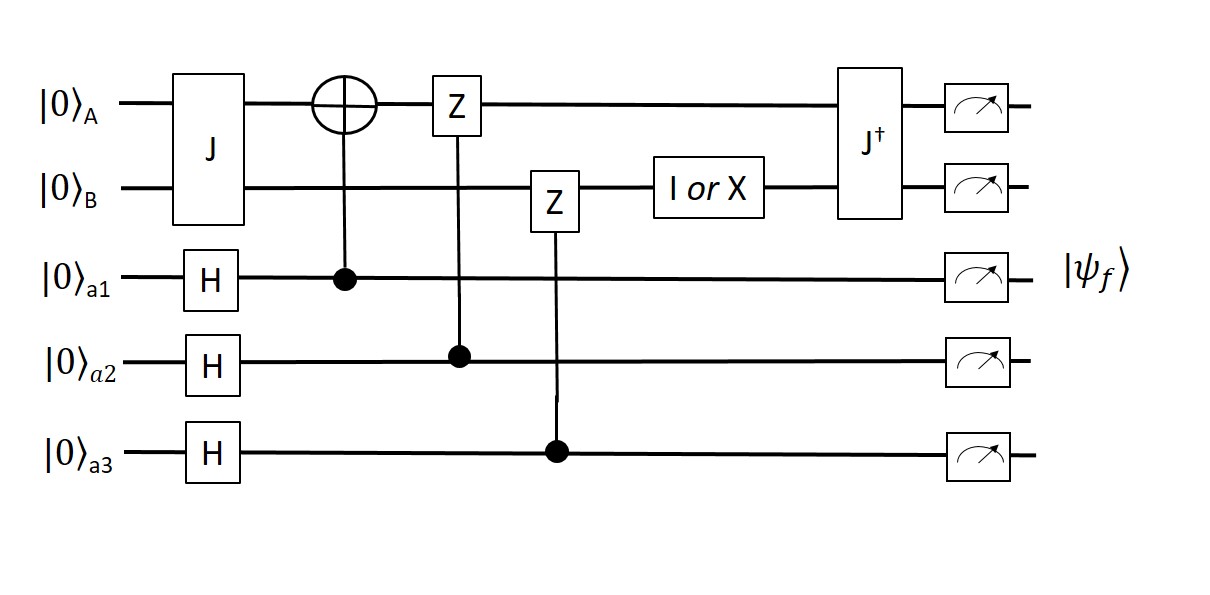}
\caption{\label{fig:implementation} Circuit used for parallel implementation of two-player quantum games. The top two qubits correspond to the two players, and the bottom three qubits are auxillory qubits used to run various strategy choices in parallel. The two player qubits are entangled, $J(\chi)$ and a Hadamard is applied to each of the auxiliary qubits. The CNOT and C-Z gates followed by either an $I$ or $X$ gate, entangle the player qubits with the auxiliary qubits so that all 16 strategy choice combinations can be run with two 5-qubit circuits. Finally, an unentangling gate is applied and a measurement is performed on all qubits. }
\end{figure}

Two of the qubits must be assigned to the players, while the three remaining qubits can be used to simultaneously run eight of the 16 strategy choice combinations.  The circuit for the parallel implementation is given in Fig. \ref{fig:implementation}. Starting with the three auxiliary qubits in the state $|000\rangle_{123}$, we apply a Hadamard gate $H$ on each of them. Together with the other 2 qubits, we form $|00+++\rangle_{AB123}$, where $|+\rangle = \frac{1}{\sqrt{2}} (|0\rangle + |1\rangle)$.

We perform $J(\chi)$ on qubits $A$ and $B$, resulting in the state
\be J_{AB}(\chi) |00+++\rangle_{AB123} \ee
We use the first auxiliary qubit as control to perform a controlled-NOT (CNOT) gate on qubit $A$. 
\be \frac{1}{\sqrt{2}}\sum_{x=0}^{1} X_A^x J_{AB}(\chi) |00x++\rangle_{AB123} \ee
We use the second auxiliary qubit as control to perform a controlled-Z (C-Z) gate on qubit $A$. 
\be \frac{1}{2}\sum_{x,y=0}^{1} Z_A^y X_A^x J_{AB}(\chi) |00xy+\rangle_{AB123}  \ee
Then we use the third auxiliary qubit as control to perform C-Z on qubit $B$. 
\be \frac{1}{2^{3/2}}\sum_{x,y,z=0}^{1} Z_B^z Z_A^y X_A^x J_{AB}(\chi) |00xyz\rangle_{AB123}  \ee
Finally, we apply $J(-\chi)$ on qubits $A$ and $B$, to obtain
\be \frac{1}{2^{3/2}}\sum_{x,y,z=0}^{1} J_{AB} (-\chi) Z_B^z Z_A^y X_A^x J_{AB}(\chi) |00xyz\rangle_{AB123}  \ee
and then measure all qubits.
The final state contains 8 terms corresponding to different strategies, $U_A \in \{ I, X, Y, Z \}$ and $U_B \in \{ I, Z\}$.

For the remaining strategies, we need to apply $X$ on qubit $B$ before the final step. In this case, the final state is

\be \frac{1}{2^{3/2}}\sum_{x,y,z=0}^{1} J_{AB} (-\chi) X_B Z_B^z Z_A^y X_A^x J_{AB}(\chi) |00xyz\rangle_{AB123}  \ee
Its terms correspond to the remaining eight strategies, $U_A \in \{ I, X, Y, Z \}$ and $U_B \in \{ X, Y\}$.

\begin{figure}
\includegraphics[width=1.\columnwidth]{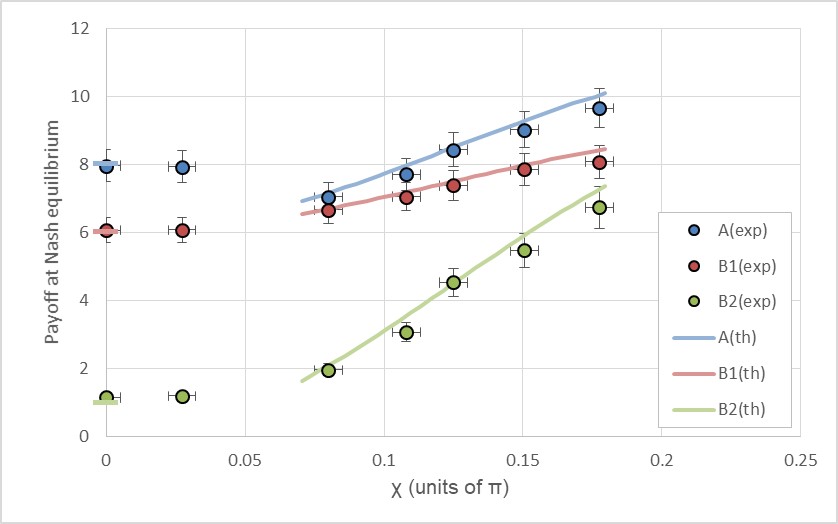}
\caption{\label{fig:bayesian}Payoffs at Nash equilibrium for the Bayesian game analyzed for p = 0.5. The outlined circles are the experimentally determined payoffs while the circles are the theoretically calculated values. There were no Nash equilibria found for regions of the graph with no data, i.e. for $\chi = 0.05$, or for $\chi >0.2$. Note also that in the case of $\chi = 0.025$ the experimental results computed a Nash equilibrium, while the theoretical computation did not.}
\end{figure}

\section{Experimental procedure}
The experimental setup constitutes a programmable quantum computer. It consists of five $^{171}$Yb$^{+}$ ions which are trapped in a linear arrangement using a radiofrequency (RF) Paul trap, and laser cooled close to their motional ground state. Two states in the $^2S_{1/2}$ ground level are used as the qubit states ($|0\rangle=\ket{F=0, m_F=0}$, and $|1\rangle= \ket{F=1, m_F=0}$). They differ in energy by the $12.642821\:$GHz hyperfine splitting which is insensitive to the magnetic field to first order. This a so-called ``atomic clock'' qubit has a typical coherence time of $~0.5\:$s which can be further extended by suppressing magnetic field noise. Optical pumping initializes the entire five-qubit register, and read-out is performed collectively by detecting state-dependent fluorescence \cite{Olmschenk07}.

 Each ion is imaged onto its own channel of a multi-channel photomultiplier tube (PMT) which allows its state to be determined with $99.4(1)\%$ average fidelity, while the $5$-qubit state detection is limited by channel-to-channel crosstalk to $95.7(1)\%$ average fidelity. For averaged state probabilities, this state-preparation and measurement (SPAM) error can be straightforwardly renormalized by applying an independently determined crosstalk-matrix. We drive qubit operations by applying a pair of Raman beams that are configured to form beat notes near the qubit frequency to the ions. Both beams are derived from a single $355\:$nm mode-locked laser. One beam is applied globally to the entire chain while the second is split into a linear array of individual addressing beams, each of which is focused onto on a single ion using a multi-channel acousto-optic modulator (AOM) \cite{Debnath16}. 

Single qubit gates (so-called R-gates) are applied by driving resonant Rabi flopping on any individual ion with the duration, phase and amplitude defined by the RF signals on the multi-channel AOM. We achieve two-qubit gates (so-called XX-gates) by applying bichromatic Raman beat notes near the motional sideband frequencies. They create an effective XX-Ising interaction between the spin degrees of freedom containing the qubit mediated by all of the collective motional modes in the ion chain \cite{Molmer99, Solano99, Milburn00}. In order to leave spin and motion disentangled at the end of the operation, we employ a pulse-shaping scheme during the gates \cite{Zhu06,Choi14}. 

Any pair of qubits can be entangled in this way, which makes this a fully-connected system of qubits  \cite{LinkePNAS2017}. A classical compiler breaks down a library of computational gates, such as Hadamard, controlled-NOT, or controlled-Phase, into the native R- and XX-operations. Since any context-dependence (such as calibration parameters) of the native operations is handled by the compiler, the high-level gates become modular. Arbitrary circuits can then be implemented from a user interface by specifying a sequence of computational and/or native gates which makes the system programmable. Native single- and two-qubit gate fidelities are typically $99.5(2)\%$ and $98.5(5)\%$, respectively. Gate times are about $10\:\mu$s for single- and $210\:\mu$s for two-qubit gates. 

\section{Results}

We perform the simulation for nine values of $\chi$ between zero and maximal in steps of $\frac{\pi}{40}$. The nominal values of $\chi$ were, in units of $\pi$: ($0$, $0.025$, $0.05$, $0.075$, $0.1$, $0.125$, $0.15$, $0.175$, $0.2$, $0.225$, $0.25$). For each of the two five-qubit cirucits depicted in Fig. \ref{fig:circuit}, we have $\sim$ 30,000 runs. Before and after each data run (except for $\chi = (0, 0.125, 0.25)$), a run was taken to measure the value of $\chi$ to take any deviation from the nominal values into account. Such deviations are the result of calibration inaccuracy in the experiment. The average of the measured $\chi$s are used for the analysis and were found to be: ($0$, $0.027(2)$, $0.054(3)$, $0.080(4)$, $0.108(4)$, $0.125$, $0.151(5)$, $0.178(5)$, $0.201(5)$, $0.224(6)$, $0.25$). For each experimental run, the quantum computer outputs the measured value (i.e. either $\ket{0}$ or $\ket{1}$) for all five qubits.

The data are parsed into two groups representing games of $A$ vs. $B_1$ and $A$ vs. $B_2$ respectively. The value of $p$, which determines the probability that player $A$ plays with either $B_1$ or $B_2$, is chosen for a given analysis, and each data point is sorted randomly into the two categories with probability $p$.  Next, the expectation value of all 32 possible outcomes is computed to form the output population vector for the two data sets. SPAM correction is then applied to the population vector to correct for readout errors.

Next, in order to compute the payoff, Eq.\ \eqref{eq:payoffB}, for the players for a given set of strategy choices, we must determine the outcomes of the qubits for player A and B depending on the 8 possible outcomes for the three auxiliary qubits. For example, if the '$I$' circuit in Fig.\ \ref{fig:implementation} outputs $\ket{000}$ for the three auxiliary qubits, this corresponds to players $A$ and $B$ having applied the strategies $\{I,I\}$. The components with the auxiliary output of $\ket{000}$ are summed to form the 4-component vector representing the expectation values of the A and B qubits for each of the 8 auxiliary output combinations. These population vectors are each then re-normalized and the experimental payoff is computed with Eq.\ \eqref{eq:payoff}. This is done for both the $I$-circuit and the $X$-circuit of Fig.\ \ref{fig:implementation} in order to compute the experimental payoff for all 16 possible strategy choice combinations of the two-player payoff matrix. The data for both types of player $B$ receive the same treatment.

From the two $4 \times 4$ two-player payoff matrices, we can compute the 64 element $4\times 4 \times 4$ payoff matrix of the Bayesian game according to Eq.\  \eqref{eq:payoffB}. The same data are analyzed for $p$  values ranging from $0$ to $1$ in steps of $0.01$ in order to observe the structure of the game as a function of the incomplete information. 

In order to compute the Nash equilibria, the best response curves must be constructed. The best response curve for player A is given by the list $\beta_A = \{ i^* , j,k\}$, where $j$ and $k$ run over all of the possible strategy choices of  $B_1$ and $B_2$, and $i^*$ is the strategy choice that gives $A$ the highest payoff for the choices of $j$ and $k$ for players $B_1$ and $B_2$. The best response curves for the $B$ players are similarly calculated,  $\beta_{B1} = \{ i , j^*,k\}$, and  $\beta_{B2}= \{ i , j,k^*\}$. The Nash equilibria are given by the intersection of the best response curves, $\beta_A \cup \beta_{B1} \cup \beta_{B2}$. In other words, any strategy choice combination where each player is playing their best response to the other player's choices is a Nash equilibrium. In general, this can result in one or more Nash equilibria or none. 

For the computation of the best response curves from the experimental data,  we consider a payoff to be the best response if it is within an amount $\delta$ of the maximum. If this is not included, the best response for each player will only be one particular strategy choice set, that will be determined by the experimental noise, and we almost never get a Nash equilibrium. The data presented use $\delta = 0.1$.

We plot the payoffs of the three players, as a function of the entanglement $\chi$, both theoretically and experimentally, for the nine different values of entanglement in Fig.\ \ref{fig:bayesian}. These data are analyzed for the case where $p = 0.5$. The error in the horizontal represents the error in entanglement calibration data taken before and after each set of data. The error in the payoff is the statistical error determined by the error in a binomial distribution of ion populations assuming that, on average, there were 3000 shots contributing to each point, with $\sim 300$ of them in $\ket{1}$, which was found to be the case for the dominant equilibria.

%%%%%%%%%%%%%%%%%%%%%%%%%%%%%%%%%%%%%%%%%%%%%%%%%%%%%%%%%%%%%%%%%%%%
The data show that for no entanglement, $\chi = 0$, the experimental data very closely match the theoretical data. With growing $\chi$ there is a systematic shift by which the experimental results fall predominantly below the theory values, probably a result of the growing error introduced by the gate. However this effect is smaller than our estimated statistical error. The theoretical and experimental data have no Nash equilibrium above a critical value of entanglement, $\chi \sim 0.175 \pi$. Note also that the experimental data for $\chi = 0.025 \pi$ show a Nash equilibrium even though none is predicted theoretically. This is indicative of the fact that experimental errors can alter the critical values of the phase-change-like behavior.  

The addition of experimental noise to quantum games is known to not affect the existence of a Nash equilibrium, but it tends to lower the payoff at Nash equilibria \cite{Buluta2006}. This is due to the fact that the quantum games that are chosen for demonstrations are those in which the payoff at Nash equilibrium is larger than the payoff for other strategy choices. This is also the case for our implementation, see Fig. \ref{fig:bayesian}. 

\begin{figure}
\includegraphics[width=1.\columnwidth]{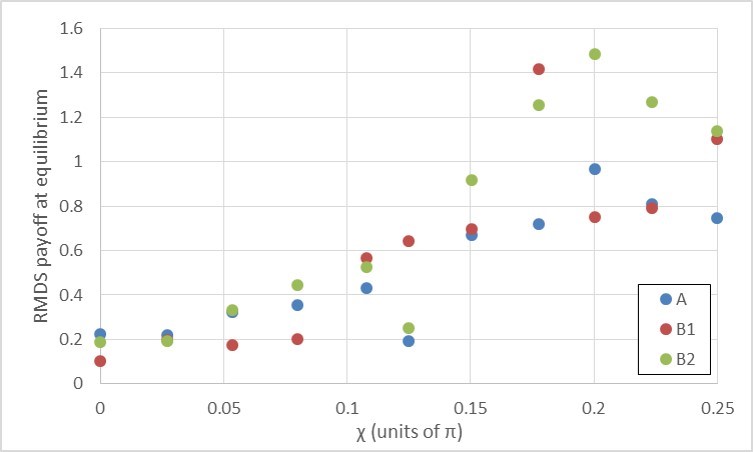}
\caption{\label{fig:RMSD} The RMSD of the payoff at the maximal payoff Nash equilibria when the Bayesian game is analyzed with $p = 0$ (i.e., player $A$ exclusively plays with player $B_2$), is plotted as a function of entanglement.}
\end{figure}

To characterize the size of the systematic shift of the expected payoff for theory vs. experiment, we plot the root-mean-square deviation (RMSD) of the experimental data to the theoretical calculation as a function of entanglement.The results are plotted in Fig. \ref{fig:RMSD}. For consistency, the data is analyzed for $p=0$, instead of $p=0.5$ because at $p=0$ there is Nash equilibrium that is common to all values of entanglement, and has the theoretical payoffs for players $(A, B_1 ,B_2)$ equal to $(11,10,9)$ with the strategy choices of the three players given by $\{ I, X, I \}$ as well as $\{Z,Y,Z\}$.  The trend is that the deviation from the theoretically calculated payoff grows as entanglement increases. Notably, for large values of entanglement, the shift is larger in the case of $p=0$ than it is for the case of $p = 0.5$ plotted in Fig. \ref{fig:bayesian}. This is because there is a larger deviation in the game of $A$ vs. $B_2$ than there is in the game of $A$ vs. $B_1$, which stems from the differences in the particular gates applied by the players in the equilibria in different regions of the Bayesian game.

\begin{figure}
\includegraphics[width=1.\columnwidth]{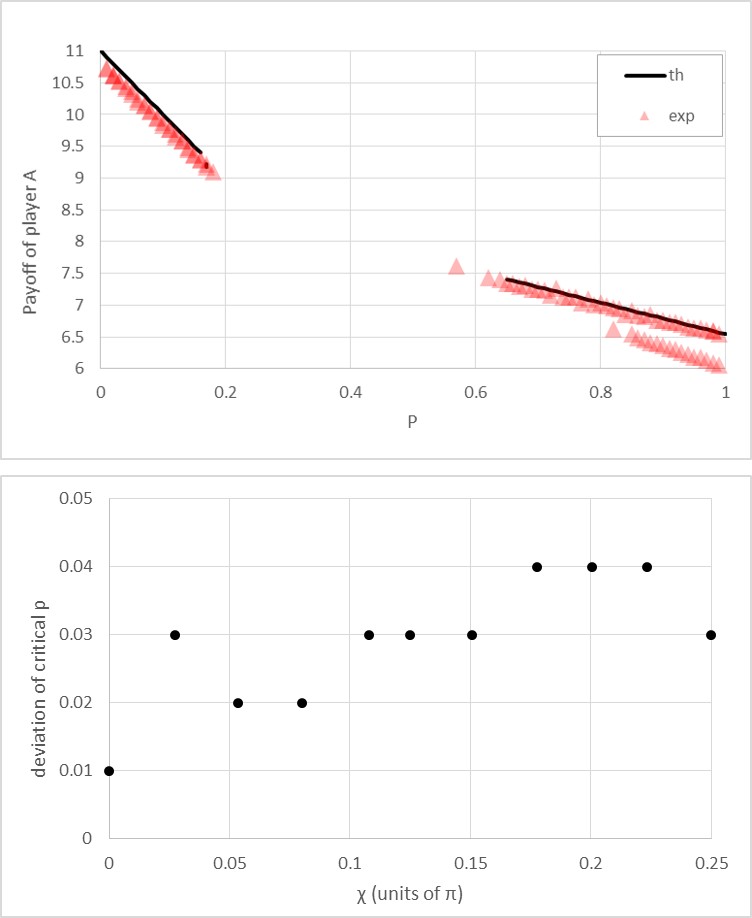}
\caption{\label{fig:EntTh} Top: example of an analysis with varying $p$ for $\chi = \pi/20$. There is a region in the center where the equilibrium disappears, and a region for high $p$ where a second equilibrium appears. Bottom: The deviation of the experimentally determined threshold $p$ as a function of entanglement.}
\end{figure}

In addition, we can see from the data how the phase-change like behavior of the Nash equilibria change with experimental noise. Our step size of $\chi = \pi/40$  does not permit a detailed study of the threshold for phase-like behavior as the entanglement is varied. But, because we incorporate the probability in the analysis, we can see the change in the threshold probability with fine steps allowing a systematic study of the threshold probability at which the equilibria change. In the top part of Fig. \ref{fig:EntTh} we plot the data for one example value of entanglement, $\pi/ 20$, as a function of $p$ both theoretically and experimentally as a black line and red triangles, respectively. 

First, there are two phase transitions, one near $p=0.16$ and the other near $p = 0.55$. It can be seen that the critical value of $p$ for the phase transition is different from theory and experiment. If we increase the value of the best-response thresholding parameter $\delta$, the experimental Nash equilibria will extend further in either direction, also, additional equilibria may arise. 

In the region of the equilibrium between $p = 0.6$ and $p=1$, another feature of note occurs. The main equilibrium in this region is given by $\{X,Y,Z\}$. Though for $p$ between $0.81$ and $1$, a second Nash equilibrium appears. This equilibrium is given by $\{Y,X,I \}$ and has a slightly lower payoff. The presence of this equilibrium is due to the finite $\delta$ parameter in the analysis of the best response. If $\delta$ were smaller, the region of the second equilibria would shrink, but so would the region with the real equilibrium. For this, and other similar observed secondary equiliria, the transition is usually blurred, meaning the second equilibrium can appear and vanish several times with increasing $p$ before it is reliably present.

We analyze the data for each value of entanglement for all values of $p$ in order to see how the deviation of the threshold $p$ changes as the entanglement grows. For consistency, we analyze the location of the phase transition near $p=0.16$. Theoretically this phase change occurs at $p=0.16$ for all values of entanglement. The deviation of the theoretical to experimental threshold $p$ also grows slightly with the entanglement as seen in the bottom of Fig. \ref{fig:EntTh}. For the values of $\chi > .175 \pi$, another equilibrium appeared in the analysis (not shown in any graph) with the strategy choices of $\{X,Z,Y\}$. This equilibrium closely resembles the main one in this region with strategy choices $\{ I, X, I \}$ and $\{Z,Y,Z\}$ with nearly the same payoffs. However, this equilibrium has a threshold $p$ which is slightly larger, $ p \sim 0.29 $. As $\chi$ increases, the payoffs, and threshold $p$ of this equilibrium converge to the main equilibrium.

\section{discussion}
Deviations in the payoff for the experimentally determined equilibria are expected due to the finite fidelity of the circuit, Fig. \ref{fig:implementation}. If there is an error in one of the player qubits, the experimentally determined payoff will be incorrect, while if there is an error in one of the auxiliary qubits, the outputs for the player qubits in that run will be misidentified. In either case this results in a different output than that predicted by the circuit. Because the game implemented is one where the payoffs at the equilibria are typically some of the higher payoffs in the game, deviation from the theoretical behavior will typically result in a lower payoff. If a game were chosen where the equilibrium was not the highest payoff in the game, i.e. if it were not Pareto optimal, then deviation from the theoretical payoff at equilibrium would tend to increase the payoff, which we see if we change the payoff matrices for the analysis in our game. Because the gate errors tend to be larger for larger values of entanglement, the deviation from theoretical behavior increases with $\chi$. 

When the experimentally determined payoff deviates significantly from the theoretical value, this can result in the disappearance or appearance of Nash equilibria. This is seen both at the boundaries between regions with different equilibria and in the appearance of new equilibria as in the bottom and top of Fig. \ref{fig:EntTh}. These transitions can be blurred, so that they are not as precisely defined as they are theoretically. This could have impact in the applications of quantum games such as mechanism design where the game is structured in order to steer the players towards certain Nash equilibria, so that the play is self-reinforcing and thus stable. If the experimental errors remove some expected equilibrium, the players would not converge their play on the strategies of the equilibrium as expected and the game could become unstable.

When an alternate competing equlibrium appears, such as in the top of Fig. \ref{fig:EntTh}, the players could become stuck on the `wrong' equilibrium. If the game dynamically changes, for instance, if $p$ changes, the players may continue to follow the lower payoff equilibrium.

We have performed what we believe is a unique demonstration of a quantum game for several reasons. It is the first experimental demonstration performed on an ion-trap-based quantum computer, and we employ a novel parallelization scheme.  The sophistication of the ion-trap quantum computer has enabled a much more extensive demonstration that has allowed us to generate enough data to demonstrate a more complicated Bayesian game than the more commonly demonstrated two player games. This is also the first time the experimental data have been solely used to solve for the Nash equilibria of the game, which allowed us to experimentally observe the behavior of the payoff as well as phase-change-like behavior  as the amount of entanglement changes. We are also the first to show explicitly how the deviation of the theory to experiment varies as a function of the entanglement. 

The value of entanglement is interpreted as being set by a referee of the game. In the case of four allowed strategy choices, maximal entanglement is not always desired from the point of view of an optimal equilibrium. In fact, the behavior of the game becomes less predictable for larger entanglement because of the increased gate errors. These considerations would have to come into any design of a game or choices by a referee in order to promote the desired behavior. 

Advances in ion-trap quantum computers underscore the reason that they are a promising platform for quantum games in particular. In addition, ion-trap-based quantum computers are a promising candidate for quantum networking, which is crucial for some quantum game applications that require the agents to be remotely located in order to be useful. The potentially long coherence time of trapped ions would also enable the quantum hardware to interact with other systems, such as humans, classical computers, sensors, etc., as may be required for quantum game applications. 

Quantum games may play an important role in the applications of quantum computers and quantum networks as they begin to become more available. We believe this demonstration brings us one step closer to that reality.

\section*{Acknowledgements}
This work was supported by the IARPA LogiQ program, the ARO MURI on Modular Quantum Circuits, the AFOSR MURI on Optimal Quantum Measurements, the ARL Center for Distributed Quantum Information, and the NSF Physics Frontier Center at JQI.


\begin{thebibliography}{99}

\bibitem {Math}von Neumann, J. and Morgenstern, O.: Theory of games and economic behavior, Princeton University Press (1944).
\bibitem{Nash1} Nash, J.: Proc. of the National Academy of Sciences, 36, 48 (1950).
\bibitem{Nash2}Nash, J.: Non-cooperative Games Annals of Mathematics, 54 (1951).
\bibitem{Econ} Shubik, M.: Game Theory Models and Methods in Political Economy, Handbook of Mathematical Economics, 1, 285 (1981). 
\bibitem{Pol} Levy, G., Razin, R.: It Takes Two: An Explanation for the Democratic Peace, Journal of the European Economic Association (2004). 
\bibitem{Bio}Axelrod, R. M., and Dion, D.: The Further Evolution of Cooperation, Science, 242 (4884): 1385 (1988). 
\bibitem{CS} Shoham, Y.: Computer science and game theory, Communications of the ACM - Designing games with a purpose, 51, 74 (2008). 

\bibitem{Meyer1999}Meyer, D.: Quantum strategies, Phys. Rev. Lett. 82, 1052-1055 (1999). 
\bibitem {Eisert1999}Eisert, J., M. Wilkens and Lewenstein, M.: Quantum games and quantum strategies, Phys. Rev. Lett. 83, 3077-3080 (1999) 

\bibitem{Maitra2015} Maitra, A., {\sl et al.}: Proposal for quantum rational secret sharing, Phys. Rev. A 92, 022305 (2015).
\bibitem{Li2009} Li, Q., He, Y., and Jiang, J.-p.: A novel clustering algorithm based on quantum games, J. Phys. A: Math. Theor. 42, 445303 (2009).
\bibitem{Zabaleta2017} Zableta, O.G., Barrang{\'u}, J. P., and Arizmendi C. M.: Quantum game application to spectrum scarcity problems, Physica A 466 (2017).

\bibitem{Du2002}
Du, J., Li, H., Xu, X., Shi, M., Wu, J., Zhou, X., and Han, R.: Experimental Realization of Quantum Games on a Quantum Computer, Phys. Rev. Lett., 88, 137902 (2002).
\bibitem{Mitra2007}
 Mitra, A., K. Sivapriya, A. Kumar: Experimental implementation of a three qubit quantum game with corrupt source using nuclear magnetic resonance quantum information processor, Journal of Magnetic Resonance 187, p 306-313 (2007.) 

\bibitem {Prevedel2007} 
Prevedel, R., Andr´e, S., Walther, P., and Zeilinger, A.: Experimental realization of a quantum game on a one-way quantum computer, New Journal of Physics 9, 205 (2007). 
\bibitem{Zhang2008}
 P. Zhang, Y.-S. Zhang, Y.-F Huang, L. Peng, C.-F. Li and G.-C. Guo: Optical realization of quantum gambling machine, EPL 82, 30002 (2008).
\bibitem{Altepeter2009}
 J. Altepeter, M. Hall, M. Medic, M. Patel, D. Meyer, and P. Kumar: Experimental realization of a multi-player quantum game, OSA/IPNRA/NLO/SL (2009).
\bibitem{Schmid2010}
 C. Schmid, A. Flitney, W. Wieczorek, N. Kiesel, H. Weinfurter, and L. Hollenberg: Experiental implementation of a four-player quantum game, New J. of Phys. Vol 12 063031 (2010).
\bibitem{Zu2012}
 C. Zu, Y -X Wang, X-Y Chang, Z-H Wei, S-Y Zhang and L-M Duan, Experimental demonstration of quantum gain in a zero-sum game, New J. of Phys, vol 14, 033002 (2012).
\bibitem{Pinheiro2013b} 
A. R. C. Pinheiro, C. Souza, D. Caetano, J. Juguenin, A. Schmidt, A. Khoury, Vector vortex implementaion of a quantum game, J. Opt. Soc. Am. B. Vol 30, no 12 (2013).
\bibitem{Balthazar2015} 
W. Balthazar, M. Passos, A. Schmidt, D. Caetano, and J. Huguenin: Experiemntal realization of the quantum duel game using linear optical circuits, Journal of Physics B: Atomic, Molecular and Optical Physics 48, 165505 (2015.)

\bibitem{Chen2006} K-Y. Chen, and T. Hogg: How well do people play a quantum prisoner's dilemma?, Quantum Information Processing, Vol 5, no 1. (2006).
\bibitem{Chen2008} K-Y. Chen, and T. Hogg: Experiments with probabilistic quantum auctions, Quantum Information Processing, Vol 7, p 139-152 (2008).
\bibitem{Blinov2004} Blinov, B. B., D. L. Moehring, L.-M. Duan, and C. Monroe, Observation of entanglement between a single trapped atom and a single photon, Nature 428, 153-157 (2004).
\bibitem{Harsanyi1967}
Harsanyi, J. C.: Games with incomplete information played by Bayesian players, Mgt. Sci. 14, 159 (1967)=

\bibitem {Brunner2013}
Brunner, N. and Linden, N.: Connection between Bell nonlocality and Bayesian game theory, Nature Communications, 4, 2057 (2013). 
\bibitem{Solmeyer2017}
N. Solmeyer, R. Dixon, and R. Balu, Characterizing the Nash equilibria of a three-player Bayesian quantum game, Quantum Inf Process, 16:146 (2017).

\bibitem{Du2003} 
Du, J., Li, H., Xu, X., Zhou, X., and Han, R.: Phase-transition-like behaviour of quantum games, J. Phys. A: Math. Gen 36 p. 6551-6562 (2003).

\bibitem{Solmeyer2017b}
N. Solmeyer, R. Balu, Proc. of SPIE Vol. 10212 102120T-1  ArXiv: 1703.03292

\bibitem{Landsburg2011}
 Landsburg, E. S.: Nash equilibria in quantum games: Proc. American Math. Soc. 139, 4423 (2011) (arXiv:1110.1351).

\bibitem{Buluta2006}
 Buluta, I. M.; Fujiwara, S.; Hasegawa, S.: Quantum games in ion traps, Physics Letters A 358, 100 (2006).

\bibitem {Debnath16}
 Debnath, S., Linke, N. M., Figgatt, C., Landsman, K. A., Wright, K. and Monroe, C., Demonstration of a small programmable quantum computer module using atomic qubits, Nature, vol. 536, pp. 63--66,(2016)  doi: 10.1038/nature18648

\bibitem{LinkePNAS2017} Linke, N. M., Maslov, D., Roetteler, M., Debnath, S., Figgatt, C., Landsman, K. A., Wright, K. and Monroe, C., Experimental comparison of two quantum computing architectures, Proc. Natl. Acad. Sci., vol. 114, no. 13, (2016).  doi: 10.1073/pnas.1618020114


\bibitem{Olmschenk07}Olmschenk, S., Younge, K. C., Moehring, D. L., Matsukevich, D. N., Maunz, P., Monroe, C., Manipulation and detection of a trapped ${\mathrm{Yb}}^{+}$ hyperfine qubit, Phys. Rev. A., vol. 76, iss 5, 052314, (2007). doi: 10.1103/PhysRevA.76.052314


\bibitem{Molmer99}M\o{}lmer, Klaus and S\o{}rensen, Anders, Multiparticle Entanglement of Hot Trapped Ions, Phys. Rev. Lett., vol. 82, iss. 9, pp. 1835-1838, (1999).  doi: 10.1103/PhysRevLett.82.1835

\bibitem{Milburn00}Milburn, G.J., Schneider, S. and James, D.F.V., Ion Trap Quantum Computing with Warm Ions, Fortschritte der Physik, vol. 48, no. 9-11, pp. 801-810, (2000) doi: 10.1002/1521-3978(200009)48:9/11<801::AID-PROP801>3.0.CO;2-1

\bibitem{Solano99}Solano, E., de Matos Filho, R. L. and Zagury, N., Deterministic Bell states and measurement of the motional state of two trapped ions, Phys. Rev. A, vol. 59, iss 4, pp R2539-2543, (1999). doi: 10.1103/PhysRevA.59.R2539

\bibitem{Zhu06}Zhu, S.-L., Monroe, C. and Duan, L.-M., Trapped Ion Quantum Computation with Transverse Phonon Modes, Phys. Rev. Lett., vol 97, iss 5, 050505, (2006). doi: 10.1103/PhysRevLett.97.050505

\bibitem{Choi14}Choi, T., Debnath, S., Manning, T. A., Figgatt, C., Gong, Z.-X., Duan, L.-M. and Monroe, C., Optimal Quantum Control of Multimode Couplings between Trapped Ion Qubits for Scalable Entanglement, Phys. Rev. Lett., vol. 112, iss 19, 190502, (2014). doi: 10.1103/PhysRevLett.112.190502


\end{thebibliography}
\end{document}